\documentclass[11pt]{article}

\usepackage{amsfonts,latexsym,amsthm,fullpage}

\def\Z{{\mathbb Z}}

\def\C{{\mathbb C}}

\newtheorem{theorem}{Theorem}
\newtheorem{lemma}[theorem]{Lemma}

\newtheorem{corollary}[theorem]{Corollary}

\begin{document}
\title{Efficient quantum algorithms for some
instances of the non-Abelian hidden
subgroup problem\thanks{
Research partially supported by 
the EU 5th framework programs QAIP IST-1999-11234, and
RAND-APX, IST-1999-14036, by
OTKA Grant No.~30132, and by an NWO-OTKA grant.
\smallskip}}
\author{
G\'abor Ivanyos\thanks{
Computer and Automation Institute,
Hungarian Academy of Sciences,
L\'agym\'anyosi u. 11.,
H-1111 Budapest,
Hungary,
e-mail:  {\tt Gabor.Ivanyos@sztaki.hu}. 
}
\and
Fr\'ed\'eric Magniez\thanks{
CNRS--LRI, UMR 8623 Universit\'e Paris--Sud, %b\^at. 490,
91405 Orsay, France,
e-mail:  {\tt magniez@lri.fr}.
}
\and
Miklos Santha\thanks{
CNRS--LRI, UMR 8623 Universit\'e Paris--Sud, %b\^at. 490,
91405 Orsay, France,
e-mail:  {\tt santha@lri.fr}.
}
}

\def\ket#1{{|{#1}\rangle}}

\maketitle

\begin{abstract}
In this paper we show that certain special cases
of the hidden subgroup problem can be solved
in polynomial time by a quantum algorithm. These
special cases involve finding hidden
normal subgroups of solvable groups 
and permutation groups,
finding hidden subgroups of groups with small
commutator subgroup and of groups
admitting an elementary Abelian normal 2-subgroup
of small index or with cyclic factor group.
\end{abstract}

\section{Introduction}

A growing trend in recent years in quantum computing is to cast
quantum algorithms in a group theoretical setting. Group theory provides
a unifying framework for several quantum algorithms,
clarifies their key ingredients, and therefore contributes to a better
understanding why they can, in some context, be more
efficient than the best known classical ones.

The most important unifying problem of group theory for the purpose
of quantum algorithms turned out to be the {\em hidden subgroup problem}
(HSP) which can be cast in the following broad terms.  Let $G$ be a finite
group (given by generators), and let $H$ be a subgroup of $G$.
We are given (by an oracle) a function $f$ mapping $G$ into a finite set
such that $f$ is constant and distinct on different left cosets of $H$,
and our task is to determine the unknown subgroup $H$.

While no classical algorithm is known to solve this problem in time faster
than polynomial in the order of the group, the biggest success of
quantum computing until now is that it can be solved by a quantum
algorithm {\em efficiently}, which means
in time polynomial in the logarithm of the
order of $G,$ whenever the group is Abelian. The main tool for this
solution is the (approximate) quantum Fourier transform which
can be efficiently implemented by a quantum algorithm \cite{Kit}.
Simon's algorithm for finding an xor-mask \cite{Sim},
Shor's seminal factorization
and discrete logarithm finding algorithms \cite{Shor}, Boneh and Lipton's
algorithm for finding hidden linear functions \cite{BoLi} are all
special cases of this general solution, as well as the algorithm of
Kitaev \cite{Kit} for the Abelian stabilizer problem, which was the first
problem set in a general group theoretical framework. That all these
problems are special cases of the HSP,
and that an efficient solution comes easily
once an efficient Fourier transform is at our disposal,
was realized and formalized by several people, including
Brassard and H{\o}yer \cite{BrHo}, Mosca and Ekert \cite{MoEk} and Jozsa
\cite{Jo1}. An excellent description of the general solution can be found
for example in Mosca's thesis \cite{Mo}.

Addressing the HSP in the non-Abelian case is considered to be
the most important challenge at present in quantum computing.
Beside its intrinsic mathematical interest, the importance of this
problem is enhanced by the fact that it contains as special case
the graph isomorphism problem.
Unfortunately, the non-Abelian HSP seems to be much more difficult
than the Abelian case, and although considerable efforts were spent
on it in the last years, only limited success can be reported.
R\"otteler and Beth \cite{RB} have presented an efficient quantum
algorithm for the wreath products
$\Z_2^k\wr \Z_2$. In the case of the dihedral groups, Ettinger and
H{\o}yer \cite{EttHoy} designed a quantum algorithm
which makes only $O( \log |G|)$ queries.
However, this
doesn't make their algorithm
efficient since the (classical) post-processing stage
of the results of the queries is done in exponential time in $O( \log |G|)$.
Actually, this result was extended by Ettinger,
H{\o}yer and Knill \cite{EHK}
in the sense that they have shown that in any group,
with only $O(\log |G|)$ queries to the oracle, 
sufficiently statistical information can be obtained to solve the
the HSP. However, it is not known how to implement efficiently
these queries, and therefore even the ``quantum part" of their algorithm
is remaining exponential.
Hallgren, Russel and Ta-Shma \cite{HRT} proved that the generic
efficient quantum procedure
for the HSP in Abelian groups works also for non-Abelian groups to
find any normal subgroup, under the condition that the Fourier transform
on the group can efficiently be computed. Grigni, Schulman, Vazirani 
and Vazirani could show that the HSP is solvable efficiently in groups
where the intersection of the normalizers of all subgroups is
large \cite{GSVV}.
A recent survey on the status of the non-Abelian HSP problem was
realized by Jozsa \cite{Jo2}.

In a somewhat different line of research, recently
several group theoretical problems have been considered 
 in the context of
black-box groups.
The notion of {\em black-box groups} has been
introduced 
by Babai and Szemer\'edi
in \cite{BabSzem}.
In this model, the elements of a group $G$
are encoded by words over a finite alphabet,
and the group operations are performed
by an oracle (the black box). The groups are assumed
to be input by generators, and the encoding is not necessarily unique.
There has been a considerable effort
to develop classical algorithms for computations with
them \cite{BeaBab,BabBea,KS},
for example to identify the composition factors
(especially the non-commutative ones).
%The size of the input
%is $n$ times the number of generators.
Efficient black-box algorithms give rise automatically to efficient
algorithms whenever the black-box operations can 
be replaced by efficient procedures.
Permutation
groups, matrix groups over finite fields and even
finite matrix groups over algebraic number fields
fit in this model. 
In particular, Watrous \cite{Wat} has recently considered solvable black-box
groups in the restricted model
of unique encoding,
and using some new quantum algorithmical ideas, he could
construct efficient quantum algorithms for
finding composition series,
decomposing
Abelian factors,
computing the order 
and testing membership in these groups.

In this paper we will focus on the HSP, and we will show that it
can be solved in polynomial time in several black-box groups.
In particular, we will present efficient quantum algorithms for
this problem for groups with small
commutator subgroup and for groups
having an elementary Abelian normal 2-subgroup
of small index or with cyclic factor group.
Our basic ingredient will be a series of
deep algorithmical results of Beals and Babai from classical
computational group theory. Indeed, in
\cite{BeaBab} they have shown that,
up to certain computationally difficult subtasks
-- the so-called Abelian obstacles --
such as factoring integers and constructive
membership test in Abelian groups
%(a generalization
%of taking the discrete logarithms),
many
problems related to the structure of black-box groups,
such as finding composition series, can be solved
efficiently for groups without large composition
factors of Lie type, and in particular, for solvable groups.
As quantum computers can factor integers and take discrete
logarithms,
and, more generally, perform the constructive
membership test in Abelian groups efficiently,
one expects that a large
part of the Beals--Babai algorithms can be
efficiently implemented by quantum algorithms.
Indeed, the above results of Watrous partly fulfill
this task, although his algorithms are not using the Beals--Babai
algorithms.
Here we will describe efficient
quantum implementations of some
of the Beals--Babai algorithms.
%(mainly for black-box groups with
%unique encoding)
%because knowledge of the structure of the
%underlying group appears to be very
%useful in the existing attempts to
%solving special cases of the hidden subgroup
%problem.
It turns out, that beside paving the way for solving the HSP in the
groups mentioned previously,
these
implementations give also 
almost ``for free"
efficient solutions
for finding hidden {\em normal}
subgroups in many cases,
including solvable groups and permutation groups.

The rest of the paper is structured as follows.
In Section 2 we review the necessary definitions about black-box groups
in the quantum computing framework, and will
summarize the most important results about Abelian and solvable
groups. In Section 3 we state the result of Beals and Babai
and {\bf Corollary \ref{BBalgquant}} which makes explicit two
hypotheses (disposability of oracles for order computing and for
constructive membership test in elementary Abelian subgroups)
under which the algorithms have efficient quantum implementations.
Section 4 deals with these quantum implementations in the following
cases: unique encoding ({\bf Theorem \ref{unique}}), modulo a hidden
normal subgroup ({\bf Theorem \ref{normal}}) and
modulo a normal subgroup given by generators in case of unique encoding 
({\bf Theorem \ref{theoter}}). As a consequence, we can derive the 
efficient quantum solution for the normal HSP in solvable and 
permutation groups
{\em without any assumption on computability of
noncommutative Fourier transforms}
({\bf Theorem \ref{theogenerator}}). 
Section 5 contains the efficient
algorithm for the HSP for groups with small commutator subgroup
({\bf Theorem \ref{commutator}}), and Section 6 for groups having
an elementary Abelian normal 2-subgroup
of small index or with cyclic factor group ({\bf Theorem \ref{elementary}}).

\section{Preliminaries}

In order to achieve sufficiently general results
we shall work in the context of black-box groups.
We will suppose that the elements of the group $G$
are encoded by binary strings of length $n$ for
some  fixed integer $n$, what we call the {\em encoding length}.
The groups will be given by
generators, and therefore
the {\em input size} of a group is
the product of the encoding length and the number of generators.
Note that the encoding of group elements need
not to be unique, a single group element
may be represented by several strings. If
the encoding is not unique, one also needs
an oracle for identity tests. Typical
examples of groups which fit in this model are
factor groups $G/N$ of matrix groups $G,$ where
$N$ is a normal subgroup of $G$ such that
testing elements of $G$ for membership in $N$ 
can be accomplished efficiently. Also, every binary string of length
$n$ does not necessarily corresponds to a group element. If the
black box is fed such a string, its behavior can be arbitrary on it.

Since we will deal with black-box groups
we shall shortly describe them in the framework of quantum
computing (see also \cite{Mo} or \cite{Wat}).
For a general introduction to quantum 
computing the reader might consult \cite{Gr} or \cite{NiCh}.
We will work in the quantum Turing machine model.
For a group $G$ of encoding length $n,$ the black-box will be given
by two oracles $U_G$ and its inverse $U_G^{-1}, $ both
operating on $2n$ qubits. For any group elements
$g,h \in G,$ the effect of the oracles is the following:
$$U_G\ket{g}\ket{h} = \ket{g}\ket{gh},$$
and
$$U_G^{-1}\ket{g}\ket{h} = \ket{g}\ket{g^{-1}h}.$$
The quantum
algorithms we consider might make errors, but the probability
of making an error should be bounded by some fixed constant
$0 < \varepsilon < 1/2.$

Let us quote here two basic results about quantum group algorithms
respectively in Abelian and in solvable black-box groups.

\begin{theorem}[Cheung and Mosca \cite{ChMo}]
Assume that $G$ is an Abelian black-box group with unique encoding. 
Then the decomposition of $G$ into a direct sum of cyclic groups of 
prime power order can be computed in time polynomial in the input 
size by a quantum algorithm.
\end{theorem}

\begin{theorem}[Watrous \cite{Wat}]\label{wat}
Assume that $G$ is a solvable black-box group with unique encoding.
Then computing 
the order of $G$ and testing membership in $G$
can be solved in time polynomial in the input size 
by a quantum algorithm. Moreover, it is possible to
produce a quantum state that approximates the pure state
$\ket{G}=|G|^{-1/2}\sum_{g\in G}\ket{g}$ with accuracy $\varepsilon$
(in the trace norm metric) in time polynomial in the 
$\mbox{input size}+\log(1/\varepsilon)$.
\end{theorem}

When we address the HSP, we will suppose that
a function $f:\{0,1\}^n \rightarrow \{0,1\}^m$ is given
by an oracle, such that for some subgroup $H\leq G$ 
the function $f$ is constant on the left cosets
of $H$ and takes different values on different cosets.
We will say that $f$ {\em hides} the subgroup $H.$
The goal is to find generators for $H$
in time polynomial in the size of $G$ and $m$, that is
we assume that $m$ is also part of the input in unary.
The following theorem resumes the status of this problem
when the group is Abelian.

\begin{theorem}[Mosca \cite{Mo}]\label{mo}
Assume that $G$ is an Abelian black-box group with %not necessarily
unique encoding. Then the hidden subgroup 
problem can be solved in time polynomial in the input size 
by a quantum algorithm.
\end{theorem}

\section{Group algorithms}
%for the structure of groups
%and hidden normal subgroups}

\label{structsect}

In \cite{BeaBab} Beals and Babai described probabilistic Las
Vegas algorithms for several important tasks
related the structure of finite black-box groups.
In order to state their result, we will need
some definitions,
in particular the definition of the parameter
$\nu (G)$, where  $G$ is any group.

Let us recall that
a {\em composition series} of a group $G$ is a sequence of subgroups
$G=G_1\rhd G_2\rhd\ldots\rhd G_t=1$ such that each $G_{i+1}$ is
a proper normal subgroup in $G_i,$ and the factor groups $G_i/G_{i+1}$
are simple.
The factors $G_i/G_{i+1}$ are
called the {\em composition factors} of $G.$
It is known that the
composition factors of $G$ are -- up to
order, but counted with multiplicities -- uniquely
determined by $G$. Beals and Babai define the parameter $\nu(G)$ 
as the smallest natural number $\nu$ such that
every non-Abelian composition factor of $G$ possesses
a faithful permutation representation
of degree at most $\nu$.

By definition, for a solvable group $G$
the parameter $\nu(G)$ equals $1$. Also,
representation-theoretic results of
\cite{FeitTits} and  \cite{LandazuriSeitz} imply
that $\nu(G)$ is polynomially bounded
in the input size in many important special cases,
such as permutation groups or even finite matrix groups
over algebraic number fields.

The {\em constructive membership test in Abelian subgroups}
is the following problem. Given pairwise commuting
group elements $h_1,\ldots,h_r,g$ of a non necessarily
commutative group, either express
$g$ as a product of powers of the $h_i$'s or report
that no such expression exists. Babai and Szemer\'edi
have shown in \cite{BabSzem} that
under some group operations oracle this problem cannot
be solved in polynomial time by classical algorithms.
This test is usually required only for {\em elementary
Abelian groups}, that is groups which are
isomorphic to $\Z_p^n$ for some prime $p$ and integer $n$.

%Under the assumption that a superset of
%the primes dividing the order of the 
%black-box group $G$ is given, the algorithms run 
%in time polynomial in the $\mbox{\it input size}+\nu(G)$ 
%and make calls to oracles for computing
%discrete logarithms in (the multiplicative
%groups of) finite fields and for 
%constructive membership tests in elementary  
%Abelian subgroups of the underlying group.

We can now quote part of the main results of \cite{BeaBab}.

\begin{theorem}
\label{BBalg}
{\bf (Beals and Babai \cite{BeaBab}, Theorem 1.2)}
Let $G$ be a finite black-box group with not necessarily
unique encoding. Assume that the
followings are given:
\begin{enumerate}
\item[(a)] a superset of the primes dividing the order of $G$,
\item[(b)] an oracle for taking discrete logarithms in finite
fields of size at most $|G|$,
\item[(c)] an oracle for the constructive membership
tests in elementary Abelian subgroups of $G$.
\end{enumerate}
Then the following tasks can be solved by
Las Vegas algorithms of running time polynomial in
the $\mbox{input size}+\nu(G)$:
\begin{enumerate}
\item[(i)] test membership in $G$,
\item[(ii)] compute the order of $G$ and a presentation for $G$,
\item[(iii)] find generators for the center of $G$,
\item[(iv)] construct a 
composition series $G=G_1\rhd G_2 \rhd\ldots\rhd G_t=1$ for $G$,
together with nice representations of the composition factors
$G_i/G_{i+1}$,
\item[(v)] find Sylow subgroups of $G$.
\end{enumerate}
\end{theorem}

A {\em presentation} of $G$ 
is a sequence $g_1,\ldots,g_s$ of generator elements for $G$,
together with a set of group expressions
in variables $x_1,\ldots,x_s$, called the
{\em relators}, such that $g_1,\ldots,g_s$
generate $G$ and the kernel of the
homomorphism from the free group
$F(x_1,\ldots,x_s)$ onto $G$ sending
$x_i$ to $g_i$ is the smallest normal
subgroup of $F(x_1,\ldots,x_s)$ containing the relators.
We remark that the generators
in the presentation may
differ from the original generators
of $G$.

A {\em nice representation} of a factor
$G_{i}/G_{i+1}$ means a homomorphism from $G_{i}$
with kernel $G_{i+1}$ to either a permutation group
of degree polynomially bounded in
the $\mbox{input size}+\nu(G)$ 
or to $\Z_p$ where $p$ is a prime  
dividing $|G|$. Of course,
if $G$ is solvable
one can insist that the representations of
all the cyclic factors be of the second kind.

It turns out that for some of the tasks
in the hypotheses of 
Theorem~\ref{BBalg} there are
efficient quantum algorithms.
By Shor's results \cite{Shor}, the oracle for
computing discrete logarithms can be implemented
by a polynomial time quantum algorithm. 
Also, a superset of the primes dividing $|G|$ can be obtained
in polynomial time by quantum algorithms in the most natural cases.
For example, if $G$ is a matrix group over a finite field,
say $G\leq \mbox{GL}(n,q)$ then such a superset
can be obtained by factoring the number 
$(q^n-1)(q^n-q)\cdots(q^n-q^{n-1})$,
the order of the group ${GL}(n,q)$. The same
method works even for factors of matrix groups
over finite fields. If $G$ is (a factor of)
a finite matrix group of characteristic zero, 
then the situation is even better because in
that case the prime divisors of $G$ are
of polynomial size. But in any case,
one can note that the superset of 
the primes dividing the order of $G$ is only used in 
Theorem~\ref{BBalg} to compute (and factorize) the orders of
elements of $G$ as well as those of matrices
over finite fields of size at most $|G|$.
This latter task can also be achieved by a
quantum algorithm in polynomial time.

In addition, we remark that the algorithm for
testing membership can be understood in 
a stronger, {\em constructive} sense,
(see  Section~5.3 in \cite{Bea}), which is the proper generalization
of the constructive membership test in the Abelian case. 
For this we need the notion of a
{\em straight line program} on 
a set of generators. This is
a sequence of expressions $e_1,\ldots,e_s$ where
each $e_i$ is either of the form
$x_i:=h$ where $h$ is a member of
the generating set or of the form
$x_i=x_jx_k^{-1}$ where $0<j,k<i$.
It turns out that for elements $g$ of $G$ one can 
also require that a straight line program
expressing $g$ in terms of the generators be
returned. 
%
\iffalse
Once the nice representation of the composition
factors have been constructed this can be
done in polynomial time
in the $\mbox{input size} + \nu(G)$) as follows.
Let $\phi_1$ be the nice representation of
$G_1/G_2$. Then we find a straight line
program expressing the image of $g$ under
$\phi_1$ in terms of the images of the generators
by solving the constructive membership
problem in the image of $G_1$ 
either by the standard permutation group algorithms
in time $\nu(G_1/G_2)^{O(1)}$
(if $\phi_1$ is a permutation representation)
or by elementary number theoretic computations
in time $(\log p)^{O(1)}$ (if $\phi_1(G_1)=\Z_p$).
Then we compute the quotient of $g$ by
the element obtained by evaluating the straight 
line program on the generators. In this way
we reduce the task to a similar problem in
$G_2$. Note that in order to compute 
the image $\phi_1(g)$ we may need to solve
the discrete logarithm problem which is feasible in quantum polynomial time.
\fi
%
Therefore, one can immediately derive from Theorem~\ref{BBalg}
the following result.

\begin{corollary}
\label{BBalgquant}
Let $G$ be a finite black-box group with not necessarily
unique encoding. Assume that the
following are given:
\begin{enumerate}
\item[(a)] an oracle for computing the orders of elements of $G$,
\item[(b)] an oracle for the constructive membership
tests in elementary Abelian subgroups of $G$.
\end{enumerate}
Then the following tasks %listed in Theorem~\ref{BBalg} and
%the following task 
can be solved by
{\em quantum} algorithms of running time polynomial in the
$\mbox{input size}+\nu(G)$:
\begin{enumerate}
%\item[(i')] constructive membership test in $G$,
\item[(i)] constructive membership test in $G$,
\item[(ii)--(v)] as in Theorem~\ref{BBalg}.
%\item[(i)] test membership in $G$,
%\item[(ii)] compute the order of $G$ and a presentation for $G$,
%\item[(iii)] find (generators for) the center of $G$,
%\item[(iv)] construct a 
%composition series $G=G_1\rhd G_2 \rhd\ldots\rhd G_t=1$ for $G$,
%together with nice representations of the composition factors
%$G_i/G_{i+1}$,
%\item[(v)] find Sylow subgroups of $G$.
\end{enumerate}
\end{corollary}

\iffalse
Once the nice representation of the composition
factors have been constructed this can be
done in polynomial time
in the $\mbox{input size} + \nu(G)$) as follows.
Let $\phi_1$ be the nice representation of
$G_1/G_2$. Then we find a straight line
program expressing the image of $g$ under
$\phi_1$ in terms of the images of the generators
by solving the constructive membership
problem in the image of $G_1$ 
either by the standard permutation group algorithms
in time $\nu(G_1/G_2)^{O(1)}$
(if $\phi_1$ is a permutation representation)
or by elementary number theoretic computations
in time $(\log p)^{O(1)}$ (if $\phi_1(G_1)=\Z_p$).
Then we compute the quotient of $g$ by
the element obtained by evaluating the straight 
line program on the generators. In this way
we reduce the task to a similar problem in
$G_2$. Note that in order to compute 
the image $\phi_1(g)$ we may need to solve
the discrete logarithm problem which is feasible in quantum polynomial time.
\fi

\section{Quantum implementations}
In this section we will discuss several cases when the 
remaining tasks in the
hypotheses of Corollary~\ref{BBalgquant} can also be efficiently
implemented by quantum algorithms.

\subsection{Unique encoding}

If we have a unique encoding for the elements of  
the black-box group $G$ then we can use Shor's order finding method.
As we will show,
in that case there is also an efficient
quantum algorithm for the constructive membership
test in elementary (and non-elementary) Abelian subgroups.
Therefore we  %use Corollary~\ref{BBalgquant}
will get the following result.
\begin{theorem}\label{unique}
Assume that $G$ is a black-box group with unique 
encoding. Then,
each of the tasks listed in Corollary~\ref{BBalgquant}
can be solved in time polynomial 
in the $\mbox{input size} + \nu(G)$ by a quantum
algorithm..
\end{theorem}
\begin{proof}
Let us prove that task {\it (b)}
in Corollary~\ref{BBalgquant} can be solved efficiently
by a quantum algorithm.
In fact, we can reduce the test to an instance of 
the Abelian hidden subgroup problem as follows. 
First, we compute the orders of the underlying elements
(see~\cite{Mo} for example). 
Let the orders of $h_1,\ldots,h_r$
and $g$ be $s_1,\ldots,s_r$ and $s$, respectively. Then
for a tuple $(\alpha_1,\ldots,\alpha_r,\alpha)$ from
$\Z_{s_1}\times\cdots\times\Z_{s_r}\times\Z_s$, set
$\phi(\alpha_1,\ldots,\alpha_r,\alpha)=
h_1^{\alpha_1}\cdots h_r^{\alpha_r}g^{-\alpha}$.
Clearly $\phi$ is a homomorphism from 
$\Z_{s_1}\times\cdots\times\Z_{s_r}\times\Z_s$
into $G$, therefore this is an instance of the Abelian hidden subgroup 
problem, and its kernel can be found in polynomial time by a quantum 
algorithm.
The kernel %of $\phi$ 
contains an element the last coordinate of which is relatively
prime to $s$ if and only if $g$ is representable as
a product of powers of $h_i$'s. Also, from such an element
an expression for $g$ in the desired form can be
constructed efficiently. % using elementary number theory.
%We have obtained the following result.
\end{proof}

This result generalizes the order finding algorithm of Watrous 
(Theorem~\ref{wat} in \cite{Wat}) 
for solvable groups.
Also note that, even if $G$ is solvable, the way how quantum 
algorithms are used here is slightly different from that of Watrous.

\subsection{Hidden normal subgroup}

Assume now that $G$ is a black-box group
with an encoding which is not necessarily unique,
and $N$ is a normal subgroup of $G$ given
as a hidden subgroup via the function $f.$
%i.e., by a function
%$f$ which is constant on the cosets of
%$N$ and takes distinct values on distinct cosets.
We use the encoding of $G$
for that of $G/N$. The function $f$ gives us a secondary encoding for
the elements of $G/N$.
%which is now unique in a sense. 
Although we do not have a machinery to multiply
elements in the secondary encoding, Shor's order-finding algorithm 
and even the treatment of the constructive
membership test outlined above are still applicable. 

\begin{theorem}\label{normal}
Assume that $G$ is a black-box group with not necessarily
unique encoding. Suppose that $N$ is a normal subgroup 
given as a hidden subgroup of $G$.
Then all the tasks listed in Corollary~\ref{BBalgquant}
for $G/N$ can be solved by quantum algorithms
in time polynomial in the $\mbox{input size}+\nu(G/N)$.
\end{theorem}
\begin{proof}
The proof is similar to the one of Theorem~\ref{unique},
where $\phi(\alpha_1,\ldots,\alpha_r,\alpha)=
f(h_1^{\alpha_1}\cdots h_r^{\alpha_r}g^{-\alpha})$
is taken.
%Thus we can solve the tasks listed in Theorem~\ref{BBalg}
%for $G/N$ in "polynomial" time.
\end{proof}

Let us now turn back to the original hidden subgroup problem.
We are able to solve it completely when the hidden subgroup is normal. 
Note that Hallgren Russell and Ta-Shma
\cite{HRT} have already given a solution for that case 
under the condition that one 
can efficiently construct the
quantum Fourier transform on $G$. The algorithm
presented here does not require such a hypothesis.
\begin{theorem}\label{theogenerator}
Assume that $G$ is a black-box group 
with not necessarily unique 
encoding. Suppose that $N$ is a normal subgroup 
given as a hidden subgroup of $G$. Then
generators for $N$ can be found by a
quantum algorithm in time polynomial 
in the $\mbox{input size} + \nu(G/N)$.
In particular, we can find hidden
normal subgroups of solvable black-box
groups and permutation groups in
polynomial time.
\end{theorem}
\begin{proof}
We use the presentation of $G/N$ obtained 
by the 
%the Beals--Babai 
algorithm of Theorem~\ref{normal}
to find generators for $N$.
Let $T$ be the generating set from the presentation.
If $T$ generates $G$ then it is easy to find generators for $N$.
Let $R_0$ denote the set of elements obtained
by substituting the generators in $T$ into the relators,
and let $N_{0}$ stand for the normal closure
(the smallest normal subgroup containing) of $R_{0}$.
Then $N=N_{0}$ since $N_{0}\leq N$ and
$G/N_{0}=G/N$ by definition of $T$ and $R_{0}$.

Still some care has to be taken since
%here because the generating set
%$T$ in the presentation need not coincide
%with the input generators $S$ for $G$ and
it is possible that $T$ generates
$G$ only modulo $N$, that is it might generate 
a proper subgroup of $G.$ 
Therefore some additional elements should be added to $R_{0}$.
Let $S$ be the generating set for $G$.
Using the constructive membership test for $G/N,$ we express
the original generators from $S$ modulo $N$
with straight line programs in terms 
of the elements of $T$. For each element $x\in S$ we form
the quotient $y^{-1}x$ where $y$ is
the element obtained by substituting
the generators from $T$ into the
straight line program for $x$ modulo
$N$. Let $S_0$ be the set of all the
quotients formed this way.
Note that $T$ and $S_{0}$ generate together $G$.
% and let
%$R_0$ denote the set of elements obtained
%by substituting the generators in $T$
%into the relators. It is easy to see
Then one can verify
that the normal closure of 
$R_0\cup S_0$ in $G$ is $N$.
\iffalse
(Indeed, let $N_0$ stand for the
normal closure of $R_0\cup S_0$.
We have $N_0\leq N$ since 
$R_0\cup S_0\subseteq N$. To see
the reverse inclusion, let $G_0$
stand for the subgroup of $G$
generated by $T$ and $N_0$. On one hand,
it is immediate that $G=G_0$
as we can express all the generators $S$
in terms of $T$ and $S_0$. On the other
hand, the presence of the elements $R_0$
in $N_0$ assures that $G_0/N_0=G/N_0$ is
is isomorphic to a factor group of
$G/N$. In particular, $|G/N_0|\leq |G/N|$.
Together with the inclusion $N_0\leq N$ this 
is only possible if $N=N_0$.)
\fi

Thus, from $R_0$ and $S_0$
we can find generators for $N$ in time polynomial
in the $\mbox{input size} + \nu (G/N)$
using the normal closure algorithm of 
\cite{BCFLS}. We obtained the desired result.
\end{proof}

\subsection{Unique encoding
and solvable normal subgroup}

We conclude this section with
some results obtained as
combination of the ideas presented
above with those of Watrous described
in \cite{Wat}. Assume that the encoding of the elements of $G$
is unique and a normal solvable subgroup $N$ of $G$ is given by generators.
%which is either $N$ solvable or of polynomial size.
We use the encoding of $G$ for that of $G/N$. The identity test
in $G/N$ can be implemented by an efficient quantum algorithm for testing
membership in $N$ due to Watrous (Theorem~\ref{wat}).
We
%further suppose that we
are also able
to produce (several copies of) the
uniform superposition 
$\ket{N}={1\over \sqrt{|N|}}\sum_{x\in N}\ket{x}$
efficiently.
%Obviously, this can be done
%in time $O(|N|)$, which is efficient for small subgroups $N$. 
For solvable 
subgroups $N$, we can again apply the result
of Watrous (Theorem~\ref{wat})
to produce $\ket{N}$ in polynomial time.
We will now show that
having sufficiently many copies of $\ket{N}$ at hand,
we can use ideas of Watrous for
computing orders of elements of $G/N$ and
even for performing the constructive membership test
in Abelian subgroups of $G/N$. Thus, we will have an
efficient quantum implementation of
the Beals-Babai algorithms for $G/N$.
We will first state a lemma which says that we can efficiently
solve the HSP in an Abelian group if we have an appropriate
quantum oracle.

\begin{lemma}\label{quanthidden}
Let $A$ be an Abelian group, and let $X$ be a finite set. 
Let $H\leq A$, and
let $f:A\rightarrow \C^{X}$ (given by an oracle)
such that:
\begin{enumerate}
\item For every $g\in A$, $\ket{f(g)}$ is a unit vector,
\item $f$ is constant on the left cosets of $H,$ and 
maps elements from different cosets
into orthogonal states.
\end{enumerate}
Then there exists a polynomial time quantum algorithm 
for finding the hidden subgroup $H$.
\end{lemma}

\begin{proof} %[Proof of Lemma~\ref{quanthidden}]
First we extend naturally $f$ to $G/H$: on a coset of $H$, it takes
the value $f(h)$ for an arbitrary member $h$ of the coset.
The algorithm is the standard quantum algorithm for the Abelian
hidden subgroup problem.
We repeat several times the following steps to
find a set of generators for $H$.
%\begin{description}
\begin{itemize}
\item Prepare the initial superposition: $\ket{1_G}\ket{0^m}$.
\item Apply the Abelian quantum Fourier transform in $A$ on the first 
register: $\sum_{g\in A}\ket{g}\ket{0^m}$.
\item Call $f$: $\sum_{g\in A}\ket{g}\ket{f(g)}$.
\item Apply again the Fourier transform in $A$:
$\sum_{g\in A/H,h\in H^{\perp}}\chi_{h}(g)\ket{h}\ket{f(g)}$.
\item Observe the first register.
\end{itemize}
%\item Repeat several times the previous steps and
%find a set of generators for $H$.
%\end{description}
By hypothesis, the states $\ket{f(g)}$ are orthogonal
for distinct $g\in A/H$, therefore 
an observation of the first register will give
a uniform probability distribution on $H^{\perp}$. 
After sufficient number of iterations, this
will give a set of generators for $H^{\perp}$,
which leads then to a set of generators for $H$.

Note that in the above steps it is sufficient to compute
only the approximate quantum Fourier 
transform on $A$ which can be done in polynomial time.
\end{proof}

% \begin{lemma}\label{quanthiddenapprox}
% 
% \end{lemma}

\begin{theorem}\label{theoter}
Assume that $G$ is a black-box group with
a unique encoding of group elements. Suppose
that $N$ is a normal subgroup given
by generators. Assume further that $N$
is either solvable or of polynomial size.
Then all the tasks listed in Corollary \ref{BBalgquant}
for $G/N$ can be solved by a quantum algorithm
in running time polynomial in the $\mbox{input size} + \nu(G/N)$.
\end{theorem}

\begin{proof} %[Proof of Theorem~\ref{theoter}]
For applying Corollary~\ref{BBalgquant}, one has to verify that we can 
perform tasks {\it (a)--(b)} of the corollary. If $N$ is of polynomial size,
it is trivial. Therefore we suppose that $N$ is solvable.
We will closely follow the approach indicated by Watrous in \cite{Wat}
for dealing with factor groups.

First, let $g\in G$. 
To compute the order of $g$ in $G/N$, we compute the period of the 
quantum function $f(k)=\ket{g^{k}N}$, where $k \in \{1, \ldots, m\}$ for
some multiple $m$ of the order.
This function can be computed 
efficiently since one can prepare the superposition
$\ket{N}$ by Theorem \ref{wat}, and for example we can
take $m$ as the order of $g$ in $G$.
Therefore by Lemma~\ref{quanthidden} one can find this period.
%simply note that a slight modified algorithm 
%works when only a upper bound of this order is given.

Second, let $g\in G$ and let $h_{1},\ldots,h_{r}\in G$
be pairwise commuting elements modulo $N$.
generating some Abelian subgroup $H\leq G/N$.
We compute the orders of the underlying elements on $G/N$
using the previous method. 
Let the orders of $h_1,\ldots,h_r$
and $g$ be $s_1,\ldots,s_r$ and $s$, respectively. Then
for a tuple $(\alpha_1,\ldots,\alpha_r,\alpha)$ from
$\Z_{s_1}\times\cdots\times\Z_{s_r}\times\Z_s$, set
$\phi(\alpha_1,\ldots,\alpha_r,\alpha)=
\ket{h_1^{\alpha_1}\cdots h_r^{\alpha_r} g^{-\alpha}N}$.
Then $\phi$ is a homomorphism from 
$\Z_{s_1}\times\cdots\times\Z_{s_r}\times\Z_s$
into $\C^{G/N}$. 
{From} Lemma~\ref{quanthidden},
the kernel of $\phi$ can be computed in polynomial time
by a quantum algorithm. Moreover it contains
an element the last coordinate of which is relatively
prime to $s$ if and only if $g$ is representable as
a product of powers of $h_i$s. Also, from such an element
an expression for $g$ in the desired form can be
constructed efficiently using elementary number theory.
\end{proof}

\section{Groups with small commutator subgroups}
Assume that $G$ is a black-box group
with unique encoding of elements,
and suppose that a subgroup $H$ is hidden by
a function $f$. Our next result states that
one can solve the HSP
in time polynomial in the $\mbox{input size} + |G'|$, where
$G'$ is the commutator subgroup of $G$. 
Let us recall the {\em commutator subgroup} is the smallest 
normal subgroup of $G$ containing the commutators
$xyx^{-1}y^{-1},$ for every $x,y\in G$.
\begin{theorem}\label{commutator}
Let $G$ be a black-box group with unique encoding
of elements. The hidden subgroup problem
in $G$ can be solved by a quantum algorithm in
time polynomial in the $\mbox{input size} + |G'|$.
\end{theorem}
\begin{proof}
Let $H$ be a hidden subgroup of $G$ defined by
the function $f$.
We start with the following observation.
If $N$ is a normal subgroup of $G$ and $H_1\leq H$
is such that $H_1\cap N=H\cap N$ and $H_1N=HN$, then
by the isomorphism theorem,
$H_1/(H\cap N)\cong H_1N/N\cong H/(H\cap N)$
which implies $H_1=H$.
We will generate such a subgroup $H_1\leq H$ for $N=G'$.

As the commutator subgroup 
$G'$ of $G$ consists of products conjugates of
commutators of the generators of $G$
we can enumerate $G'$, and therefore also $G'\cap H$,
in time polynomial in the $\mbox{input size}+|G'|$.
We consider the function $F:x\mapsto\{f(xG')\}=\{f(xg)|g\in G'\}$
which can be computed by querying $|G'|$ times 
the function $f$.

The function $F$ hides the subgroup $HG'$.
Note that $HG'$ is normal since $G/G'$ is Abelian.
Thus by Theorem~\ref{theogenerator},
we can find generators for $HG'$
by a quantum algorithm
in time polynomial in the size of the input + $|G'|$
since $\nu(G/HG')=1$, because $G/HG'$ is Abelian.
%and since $\mbox{size of $F$}=O(|G'|\cdot\mbox{size of $f$})$.

For each generator $x$ of $HG'$, we enumerate
all the elements of coset $xG'$ and select an element of $xG'\cap H$.
The cost of this step is again polynomial in the 
$\mbox{input size}+|G'|$.
We take for $H_1$ the subgroup of $G$ generated by
the selected elements and $H\cap G'$.
We get $H_1\cap G'=H\cap G'$,
and by the definition of the selected elements $H_{1}G'=HG'$.
\end{proof}

A group $G$ is an {\em extra-special $p$-group}
if its commutator subgroup $G'$ coincides with its center, $|G'|=p$, 
and moreover $G/G'$ is an elementary Abelian $p$-group.
Therefore we get the following corollary from the previous 
theorem.
\begin{corollary}
The hidden subgroup problem
in extra-special $p$-groups
can be solved by a quantum algorithm
in time polynomial in $\mbox{input size}+p$.
\end{corollary}

%\section{Groups with large center}
%
%{\bf Do we want to write down this???
%Maybe too trivial.}
%

\section{Groups with a large
elementary Abelian normal 2-subgroup}
 
Assume that $N$ is an elementary Abelian normal 2-subgroup
of a group $G,$ and it is given by generators as part 
of the input. Our aim is to solve the HSP in $G$
in the cases where $G/N$ is either small or cyclic.
Typical examples of groups of the latter type
are matrix groups over a field
of characteristic 2 of degree $k+1$
generated by a single matrix
of type {\it (a)}, where the $k\times k$ sub-matrix in the
upper left corner is invertible,
together with several matrices of type {\it (b)}:\\
\centerline{
$
\mbox{\it (a) }\left(\begin{array}{ccccc}
\ast & * & * & * & 0 \\
\ast & * & * & * & 0 \\
\ast & * & * & * & 0 \\
\ast & * & * & * & 0 \\
0 &  0 & 0 & 0 & 1 \\
\end{array}
\right),
\mbox{\it (b) }
\left(\begin{array}{ccccc}
1 &  0 & 0 & 0 & * \\
0 &  1 & 0 & 0 & * \\
0 &  0 & 1 & 0 & * \\
0 &  0 & 0 & 1 & * \\
0 &  0 & 0 & 0 & 1 \\
\end{array}
\right) .$
}
\\
Note that the class of groups of this kind
include the wreath products $\Z_2^k\wr \Z_2$
in which the hidden subgroup problem
has been shown to be solvable in polynomial time
by R\"otteler and Beth in \cite{RB}.
Based on a technique inspired by
the idea of 
Ettinger and H{\o}yer
used for the dihedral groups
in \cite{EttHoy}, we solve
the hidden subgroup problem 
in quantum polynomial time
in this more general class of groups.

\begin{theorem}\label{elementary}
Let $G$ be a black-box group with unique encoding
of elements and $N$ be a normal elementary
Abelian $2$-subgroup of $G$ given by generators.
Then the hidden subgroup problem
in $G$ can be solved by a quantum algorithm in
time polynomial in the $\mbox{input size}+|G/N|$.
If $G/N$ is cyclic then the hidden subgroup
problem can be solved in polynomial time.
\end{theorem}
\begin{proof}
Let $H$ be a subgroup of $G$ hidden by
the function $f$. 
The main line of the proof is like in Theorem~\ref{commutator}:
we will generate $H_1\leq H$
which satisfies $H_1\cap N=H\cap N$ and 
$H_1N/N=HN/N$ (or equivalently $H_1N=HN$).
Again we start the generation of $H_1$ with $H\cap N$
which can be computed in polynomial time in the input size
by Theorem~\ref{mo}
since $N$ is Abelian.
The additional generators of $H_1$ will be obtained from
a set $V\subseteq G$ which, for every
subgroup ${\overline M}\leq G/N$ (in particular, for
${\overline M}=HN/N$), contains some generator set
for ${\overline M}$. For each $z\in V$, we will verify if $zN\in HN$
(equivalently $zH\cap N\neq\emptyset$
or also $zN\cap H\neq\emptyset$), and in the positive case
we will find some $u\in N$ such that $u^{-1}z\in H$.
Both of these tasks will be reduced to the Abelian hidden
subgroup problem, and the elements of the form $u^{-1}z$ will
be the additional generators of $H_1$.

If $G/N$ is cyclic,
we use Theorem~\ref{theoter} to
find generators for the Sylow subgroups of
$G/N$ (note that $\nu(G/N)=1$). Each Sylow
will be cyclic (and unique), therefore a random element of
the Sylow $p$-subgroup will be a generator with
probability $1-1/p\geq 1/2$. Note that one can check if the
choosen element is really a generator by using the order finding
procedure of Theorem~\ref{theoter}.
Then, for each $p$
we choose a generator $x_p N$ for the Sylow $p$-subgroup
after iterating the previous random choice.
The $p$-subgroups of $G/N$ are
$\langle x_p N \rangle,\ldots,\langle x_p^{h_p} N\rangle=N/N$, 
where $p^{h_p}$ is the
order of the Sylow $p$-subgroup of $G/N$.
Let $V$ stand for
the union of the sets $\{1,x_p,\ldots,x_p^{h_p}\}$
over all primes $p$ dividing $|G/N|$.
Note that $|V|=O(\log |G/N|)$, and the cost of
constucting $V$ is polynomial in the input size.
$V$ contains a generating set for an
arbitray subgroup ${\overline M}$ of
$G/N$ because for each $p$, it contains 
a generator for the Sylow $p$-subgroup 
of ${\overline M}$ (namely $x_p^{l_p}$ where
$l_p$ is the smallest positive integer $l$
such that $x_p^lN\in {\overline M}$).

%We shall make use the fact that
%$HN/N$ is generated by all the cosets
%$zN$ where $z\in V$ such that $z\in HN$.

In the general case, let $V$ be
a complete set of coset representatives of $G/N$.
$V$ can be constructed by the following standard method.
We start with the set $V=\{1\}$.
In each round we adjoin to $V$ a representative 
$vg$ of a new coset, 
for each $v\in V$ and each generator $g$ of $G$,
if $vg\not\in wN$, for all $w\in V$.
This membership test can be achieved using a quantum
algorithm for testing membership of
$w^{-1}vg$ in the commutative group $N$. 
The procedure stops if no new element can be added.

Then, for each $z\in V\setminus\{1\}$,
we consider the function defined on $\Z_2\times N$
as follows. For every $x\in N$,
let $F(0,x)=f(x)$ and let $F(1,x)=f(xz)$. 
Obviously, for $i\in\{0,1\}$ and
$x,y\in N$,
$F(i,x)=F(i,y)$ if and only if
$y^{-1}x\in H\cap N$, while
$F(0,x)=F(1,y)$ if and only if
$y^{-1}x\in zH\cap N$. 

We claim that $zH\cap N$ is either empty
or a coset of $H\cap N$ in $N$.
Indeed, if $zH\cap N$ contains $zh$ for some $h\in H$,
then $zh(H\cap N)\subseteq zH\cap N$, and
conversely for all $h'\in H$ such that $zh'\in N$,
we have $(zh)^{-1}zh'=h^{-1}h'\in H\cap N$.
It follows that in the group $\Z_2\times N$,
$F$ hides either $\{0\}\times (H\cap N)$
or $\{0\}\times (H\cap N)\bigcup \{1\}\times u(H\cap N)$
for some $u\in zH\cap N$ depending on whether $zH\cap N$ 
is empty or not. 
Note that this set is indeed a subgroup because
$N$ is an elementary Abelian $2$-group.
We remark that $u$ is determined only modulo $H\cap N$. 

As $\Z_2\times N$ is Abelian, we can find generators 
for this hidden subgroup in quantum polynomial time. 
{From} any generator of type $(1,u)$ we obtain an element
$u^{-1}z\in zN\cap H$.
Repeating this,
we collect elements in $zN\cap H$
for each of $z\in V\setminus\{1\}$ such that $zN\cap H\neq \emptyset$. 
%,
%or, equivalently, $zN/N\in HN/N$.
Let $H_1$ be the subgroup of $G$
generated by the collected elements and by $H\cap N$.
Then by construction $H_{1}$ is a subgroup of $H$
which satisfies the claimed properties. 
\end{proof}

\end{document}